\documentclass[sigconf,authorversion,nonacm]{acmart}
\AtBeginDocument{%
  \providecommand\BibTeX{{%
    \normalfont B\kern-0.5em{\scshape i\kern-0.25em b}\kern-0.8em\TeX}}}

\copyrightyear{2022}
\acmYear{2022}
\setcopyright{acmlicensed}\acmConference[CHI '22 Extended Abstracts]{CHI Conference on Human Factors in Computing Systems Extended Abstracts}{April 29-May 5, 2022}{New Orleans, LA, USA}
\acmBooktitle{CHI Conference on Human Factors in Computing Systems Extended Abstracts (CHI '22 Extended Abstracts), April 29-May 5, 2022, New Orleans, LA, USA}
\acmPrice{15.00}
\acmDOI{10.1145/3491101.3519661}
\acmISBN{978-1-4503-9156-6/22/04}

\usepackage{xcolor}

\usepackage{multirow}
\begin{document}

\title{Finding Strategies Against Misinformation in Social Media:\\A Qualitative Study}

\author{Jacqueline Urakami}
\email{urakami.j.aa@m.titech.ac.jp}
\orcid{0000-0002-2866-0807}
\affiliation{%
  \institution{Tokyo Institute of Technology}
  \country{Japan}
}

\author{Yeongdae Kim}
\email{kim.y.ah@m.titech.ac.jp}
\affiliation{%
  \institution{Tokyo Institute of Technology}
  \country{Japan}
}

\author{Hiroki Oura}
\email{houra@rs.tus.ac.jp}
\affiliation{%
  \institution{Tokyo University of Science}
  \country{Japan}
}

\author{Katie Seaborn}
\email{seaborn.k.aa@m.titech.ac.jp}
\affiliation{%
  \institution{Tokyo Institute of Technology}
  \country{Japan}
}

\renewcommand{\shortauthors}{Urakami, et al.}

\begin{abstract}
  Misinformation spread through social media has become a fundamental challenge in modern society. Recent studies have evaluated various strategies for addressing this problem, such as by modifying social media platforms or educating people about misinformation, to varying degrees of success. Our goal is to develop a new strategy for countering misinformation: intelligent tools that encourage social media users to foster metacognitive skills "in the wild." As a first step, we conducted focus groups with social media users to discover how they can be best supported in combating misinformation. Qualitative analyses of the discussions revealed that people find it difficult to detect misinformation. Findings also indicated a need for but lack of resources to support cross-validation of information. Moreover, misinformation had a nuanced emotional impact on people. Suggestions for the design of intelligent tools that support social media users in information selection, information engagement, and emotional response management are presented.
\end{abstract}

\ccsdesc[500]{General and reference~Empirical studies}
\ccsdesc[500]{Networks~Social media networks}
\ccsdesc[300]{Human-centered computing~User studies}

\keywords{misinformation, social media, focus group, opinion discussion, emotional reactions}

\maketitle

\section{Introduction}
Misinformation is regarded as a fundamental challenge in the contemporary communication landscape of social media. Even though misinformation is not a new phenomenon, its increase in volume and the large number of people exposed to it has caused widespread alarm in recent years \cite{Flynn2017, Lazer2018}. It is therefore not surprising that how to address the issue of misinformation has received a lot of attention, not only from researchers but also from social media platforms themselves (e.g. "Birdwatch" by Twitter). 
Research has focused on understanding how misinformation spreads in social media (e.g. \cite{Allcott2019, Shin2018, Wu2016}, designing interventions for correcting misinformation (e.g. \cite{Vraga2017, Bautista2021, Walter2020}, exploring the (in)effectiveness of warnings (e.g. \cite{Mena2020, Ross2018}), or studying users reactions to fake news posts \cite{Geeng2020}. Other work has acknowledged the impact of emotion on technology use \cite{Venkatesh2000} and the mediating effect of emotions on people's behavior in social media \cite{Schreiner2021}. As yet, there is still much that we do not understand well about the relationship between emotion and how people handle misinformation on social media. Moreover, there is room to design new user-centred strategies that harness and/or extend what people already do to combat misinformation in social media. 


To this end, our goal was to uncover the role of emotion in the metacognitive strategies people un/knowingly deploy when confronted with social media content. We asked: \emph{RQ1: What strategies do people use when confronted with misinformation? RQ2: What role does emotion have in social media use?} Our contributions are fourfold. First, we offer qualitative evidence that people struggle to detect and manage misinformation.
Second, we provide evidence 
supporting the view that cognition and affective reactions are intimately entwined. Third, we offer a synthesis of the high-level metacognitive strategies we derived from participants' storied experiences with misinformation.
We end with an initial set of guidelines for developing an intelligent metacognitive tool for everyday use on social media platforms.

\section{Theoretical Background}
\subsection{Role of Emotion in Social Media}
Social media differs from traditional information media in being user-based, interactive, relationship- and community-driven, and valuing emotions over content \cite{Mislove2007}. Indeed, user emotion is a driving force on social media. So far, most of the literature has focused on deriving emotional expressions from social media content, particularly through sentiment analysis and other forms of large-scale, algorithmic approaches to content analysis \cite{AbdElJawad2018, Vashishtha2019, Yoo2018}. For example, Alduaiji and Datta \cite{Alduaiji2018} conducted a sentiment analysis on three different Twitter data sets and found that positive tweets are shared more often than negative tweets. Others have found that people’s emotional reactions to a topic can change over time \cite{Naskar2020} and be influenced by the emotions of others \cite{Ferrara2015, Kramer2014}. In short, social media can convey and evoke emotion. However, less is known about how emotional reactions to content, especially misinformation, relates to other aspects of social media engagement. \emph{Valence}--the affective quality of emotion as pleasant or unpleasant--and \emph{arousal}--the intensity of the emotion--seem to be related to how content is shared on social media. Nelson-Field et.al. \cite{NelsonField2013} analyzed user behavior in sharing video ads on Facebook. They found that videos rated with positive valence and high arousal were shared 30\% more often than videos with negative valence and low arousal. Similarly, Yu \cite{Yu2014} reported a positive relationship between brand posts with positive valence and users' sharing and liking of these posts. A small body of work (e.g., \cite{Wang2020}) has started to explore the potential relationship between emotion and \emph{cognition} as a mediator of interactions with misinformation on social media, however studies have also shown that emotion recognition is often perceived as being invasive and scarey \cite{Andalibi2020}. Cognitive ability has long been identified as a key variable in misinformation of all kinds, online and off \cite{Frenda2011}. In the past, the relationship between emotion and cognition was widely discussed and often contested in the psychological literature \cite{Dolan2002, Lazarus1982, Zajonc1984}. However, recent findings in neuroscience (e.g., \cite{Pessoa2008, Phelps2006}) support the view that emotion and cognition influence each other and are closely related. More work is needed to understand whether and how this plays out across a variety of topics subject to misinformation within social media.

\subsection{Employing Metacognitive Strategies in Social Media}
One cognitive strategy that could address susceptibility to and emotional reactions towards misinformation in social media is \emph{metacognition}. Originally coined by John Flavell \cite{Flavell1979}, metacognition is defined as \emph{cognition about cognitive phenomena} or more colloquially "thinking about thinking." It has been described as higher order thinking that involves active control over one's cognitive processes while engaged in learning \cite{Sternberg1986a, Sternbergb} and executive control involving self-monitoring and self-regulation \cite{Hennessey1999, Miller2000}. Lab and questionnaire studies have shown that supporting and encouraging the use of the metacognitive processes involved in processing information can enable social media users to detect misinformation more easily and to react appropriately to them \cite{Salovich2020}. 
In general, encouraging people to engage in the evaluation of information is a successful approach to reduce susceptibility to inaccuracies \cite{AndrewsTodd2021, Brashier2020}. However, people are often not motivated to engage in information evaluation, even when it seems obvious and reasonable \cite{Mena2020, Ross2018}. 

Social media users may be encouraged to employ metacognitive strategies 
in social media contexts with the right design scaffolds and user experience (UX) design. However, with no known applied work to date, it is not clear how best to approach this as a design or technical problem. We start by exploring the role of emotion and cognition in people's orientations and responses to social media content, especially misinformation, as a way of deriving strategies and potential design guidelines. As a first step, we conducted a focus group study with social media users to better understand how they grapple with social media content that is emotionally evocative and potentially misinformed.

\section{Methods}
We conducted an online qualitative study with social media users.
We qualitatively analyzed transcripts and quantitatively assessed questionnaire responses. Our qualitative approach focused on describing the experiences of the participants in everyday language \cite{Lambert2012, Sandelowski2000}. 
We chose the divisive and emotionally-charged topic of using masks as a preventive measure for COVID-19, 
assuming that participants would be aware of the ongoing discussions and may have already formed their own opinions about it. Furthermore, numerous recent studies have highlighted the urgency of addressing misinformation on COVID-19 in social networks (e.g. \cite{Roozenbeek2020, CuanBaltazar2020, Tasnim2020, Saltz2021}). 

\subsection{Participants}
We recruited a diverse group of participants from different countries using Facebook, word-of-mouth advertising, and our personal online networks. Fifteen participants (5 men, 10 women, 
aged 18-54)
representing six different ethnicities and being located in Japan, the UK, and Canada took part in the study. Most were non-native English speakers $(n = 12)$ but all could use English at a conversational level or higher. The study was approved by the university's ethical board.

\subsection{Questionnaires}
A pre- and post-questionnaire was administered to participants of the focus group using SurveyMonkey\footnote{\url{See https://www.surveymonkey.com}}. 
Tweets and comments underwent a rigorous rating process by three raters.

The
questionnaires were identical except that demographic questions were not included in the post-questionnaire. The questionnaire started with an explanation of the purpose of the research and requested participants’ consent. The first section contained questions about average social media platform usage on weekdays and on the weekend, based on the questions used by Walsh et al. \cite{Walsh2013}. Typical social networking activities were captured using the Social Networking Activity Intensity Scale (SNAIST) \cite{Li2016}. In the second section, participants were asked to read four tweets from Twitter and four comments from the New York Times about different opinions on wearing masks as preventive measure for COVID-19. These were presented in random order to each participant using SurveyMonkey’s randomization tool. After reading each tweet/comment, participants were asked to rate their emotional reaction according to valence and arousal (9-point Likert scales), as well as estimate their agreement and familiarity (5-point Likert scales) with the content. Emotional reactions were measured with the Tactile Self-Assessment Manikin (T-SAM) \cite{IturreguiGallardo2020}. 
The third section of the questionnaire contained demographic questions and questions about participants’ ideological leaning and degree of agreement on whether or not masks protect against COVID-19.
The questionnaire took about 20 minutes. 

\subsection{Equipment}
We used the Zoom video conferencing platform\footnote{See \url{https://zoom.us/}}. Each Zoom meeting was recorded and stored for later analyses. Zoom recordings were transcribed and double checked by a second experimenter. 

\subsection{Procedure}
Participants were assigned to one of four online focus groups based on their availability and time zone. We also aimed to maximize a large diversity of group members in terms of gender, age, and ethnicity. Participants had to complete the questionnaires by a day before and after the focus group meeting. The same protocol was used in each session. Each session was scheduled for 1 hour.
Each session started with a 10 minutes opening (greeting participants, introducing study goals, confirming basic rules for discussion), followed by a 40 minutes discussion section (Sharing own experiences with social media and misinformation / Importance of social media in everyday life / Experience with misinformation), followed by a 10 minutes Wrap up and Summary of the discussion. At the end of the session participants were informed about the procedure for the post-questionnaire.

\subsection{Data Analysis}
Descriptive statistics and correlations were generated for the quantitative data.
A thematic analysis \cite{Guest} of the transcripts was conducted by two raters who identified three thematic blocks (importance of social media, experience with misinformation, important discussion points). 
Responses were filtered according to the main questions for each block (see 3.4).
Rater agreement of filtered 
responses was 85\%. Discrepancies were resolved through discussion.


\section{Results}
\subsection{Questionnaire: General Usage of Social Media}
Questionnaire data revealed that participants were regularly using a variety of different social media platforms, such as YouTube $(n = 15)$, Instagram $(n = 11)$, Line $(n = 11)$, Twitter $(n = 8)$, WhatsApp $(n = 7)$, Facebook $(n = 6)$, and WeChat $(n = 4)$. Only one participant used TikTok. Estimated time spent on each platform differed greatly between participants. Averaging usage time across participants showed that most time was spent on YouTube on weekdays $(Md = 30 min)$ and weekends $(Md = 60 min)$, with some participants $(n = 4)$ spending an average of 2-4 hours on weekdays and weekends on the platform, probably because YouTube provides a large amount of content for entertainment. All participants used at least one type of social messenger application, such as WeChat, Line, and/or WhatsApp spending between 30-300 minutes on these apps on average weekdays and weekends. 

\subsection{Questionnaire: Response to Tweets and Comments}
The main purpose of the questionnaire was to familiarize focus group participants with conflicting information on social media and to use these experiences as a reference for discussion. Because of our small sample, the results of the questionnaire were not representative of the general population. 
Overall, participants were supportive of the argument that mask-wearing protects against COVID-19 $(M = 3.73, SD = 1.16)$.


Furthermore, a significant negative correlation between participants’ attitudes towards wearing a mask as a COVID-19 prevention measure and rejecting tweets/comments was found in the pre- $(r = .825, p < .01)$ and post-test $(r = .797, p <.01)$.



\subsection{Focus Group Discussion}
Findings from the focus group discussions are organized into five themes: 1) General social media use, 2) Responses to the content of tweets/comments in the pre-questionnaire, 3) Experience with misinformation, 4) Recognizing misinformation, and 5) Insights from the discussion. Participants’ quotes are provided in \textit{“italics”}. 

\subsubsection{General Social Media Use}
Most participants agreed that social media is an important part of everyday life. 
Social media had become especially important during the COVID-19 pandemic to overcome feelings of isolation and distance from family and friends $(n = 2)$. Participants felt that it was \textit{“very inconvenient when not available”}, and \textit{“can’t imagine to live without it”}. 
Social media was mainly used to stay in touch with family and/or friends $(n = 14)$. Three mentioned that they used it to connect with new people and two interacted with colleagues. Social media was also largely used as a tool to collect different types of information $(n = 6)$ such as \textit{“getting the latest news especially when something had just happened”}, \textit{“looking up information what is going on in the world”}, gathering political and world news or information about happenings and events in the local area, but also for \textit{“getting information about new people before meeting them”}. However, one participant acknowledged at the same time that social media is the \textit{“least trustable information source”}. 

\subsubsection{Responses to the Content of Tweets/Comments in the Pre-Questionnaire}
Participants felt that the content was representative of what is on social media right now. They noticed differences in formality of language and emotional tone, with some \textit{“pretty emotional”} and even \textit{“aggressive”}. Furthermore, participants recognized that these tweets/comments mainly voiced personal opinions with only some providing scientific evidence. Participants' initial reactions fell into two categories: either triggering an affective response or a cognitive evaluation. Reported affective responses were being \textit{“shocked”}, feeling \textit{“hurt”}, \textit{“upset”} and \textit{“surprised”} about anti-mask comments. Cognitive evaluations were made in regard to the origin of the poster: \textit{“they try to convey to you an idea, an opinion, an emotion and like you read them and you get an idea of the person [who wrote them]”}. Some evaluated the value of the information $(n = 6)$, as indicated by comments such as \textit{“I think about which is most relevant to me”} , \textit{“I chose which comment I want to hear and which not”}, and \textit{“It's like a piece of news and before I made my judgement, I would like to read the whole information”}.    

All noticed that conflicting information was represented in the tweets/comments. Notably, we received mixed answers when we asked about their impression of the balance in terms of supporting or rejecting mask-wearing. Two participants could not remember, two stated that their own viewpoint was mainly supported, four felt that the content was balanced, and another four indicated that most of the content rejected mask-wearing as a preventive measure. Participants then realized that their 
impressions of tweets/comments differed. They elaborated that their impression could be biased because of their own position: \textit{”because I think mask is important so I just feel like I just want to avoid this comment”}, \textit{“I think if it's [the comments] more negative because my perception is positive”}. Reading the tweets/comments also impacted how participants thought about the topic of mask-wearing. While just over half of participants stated that it did not change their way of thinking $(n = 9)$, five participants said that it slightly changed or made them think more about how they think about the topic: \textit{“it kind of swayed me a bit”}, \textit{“good to see the other side of the beliefs”}, \textit{“concerns some things which I never thought about before”} or \textit{“they gave me fresh points of view to some problems”}. 

\subsubsection{Experience with Misinformation}
There was a general agreement that misinformation is an increasing problem in social media. Participants stated that there is \textit{“so much misinformation, it is getting harder to filter it out”}, that they \textit{“don't trust social media to get important information”}, and there is \textit{“much misinformation even on credible sources”}. When asked how they have acted on misinformation in their past, four different patterns emerged: 1) ignoring \textit{“I perceive this information as noise”}, \textit{“there are a lot of false information, we are getting used to the false information”}, to \textit{“brush it off”}, 2) warning or educating friends and family about it \textit{“I told my father don't listen to him it's all wrong”}, \textit{“I re-posted to my friends to tell this is not true”}, 3) reporting it to the platform \textit{“I actually was like reporting my uncle's post because they were completely racist”}, \textit{“it is important to report posts of fake news”}, and 4) searching for additional information sources \textit{"should follow up on it if it is a new information”}, \textit{“you have to double check where that is coming from”}, \textit{“find a different source that's talking about the same news”}. One decisive factor in the type of action taken was whether the information was of personal concern or not. Participants stated \textit{“evaluate how important it is, if not important than brush it off”}, \textit{“if like the content is related too closely to my life then I try to check it out to different sources”}, and \textit{“if it is not personal than ignore, need to choose which info you care about”}. Participants also expressed that the large amount of misinformation makes it harder to follow up on: \textit{“getting a lot of misinformation in social media but it is getting harder to filter it out”}, \textit{“checking info is extra work, everybody can just post anything”}.

Many participants $(n = 8)$ also mentioned the negative emotional impact that encountering misinformation can have. Participants made comments such as \textit{“I am angry that others believed false information”}, \textit{“so worried because I didn't expect him to be somebody who can fall to that kind”}, \textit{“feel sad when people I respect post misinformation”}, \textit{“sometimes it's a bit shocking”}, when \textit{“I have a bad day, makes me feel upset but can't really do anything”}, \textit{“a lot of misinformation makes me worry about my family if they are OK or not”}, \textit{“feel angry if it is bad info about other person”}. Four participants emphasized that misinformation can be harmful and dangerous, creating division among people, whereas one participant noted that \textit{“I think it's really important to actually still have people that post false information on your Facebook because they're trapped in like a social bubble so you kind of actually post something that might change their mind”}.

\subsubsection{Recognizing Misinformation}
All participants had frequently encountered misinformation and 
agreed on what misinformation is. However, when reflecting on their approach to identifying misinformation, participants found it difficult to give a straightforward answer. Source credibility was mentioned as one way to make a judgment call $(n = 10)$, which also included checking the political background of the poster. Another 
was a lack of scientific evidence $(n = 2)$, or, alternatively, the presence of specific keywords that are often used in misinformation: \textit{“also there's a typical way like they use the term share it as much as you can, these statements probably 80 percent just fake news”}. Participants also acknowledged that it is often difficult to tell what is misinformation $(n = 7)$. In such cases they rely on their intuition \textit{“you just know”}, \textit{“it does not seem to make sense”}. Misinformation that mixes facts with false information was regarded as especially difficult to identify $(n = 2)$. Participants also agreed that cross-referencing social media information with other sources, such as an official agency, other news sites, and original scientific articles, is a good way to verify information $(n = 14)$. These alternative information sources also included friends, colleagues, and family members.

\subsubsection{Participants’ Insights from the Discussion}
Participants were asked to reflect on the discussion to stimulate metacognitive processes on their own perceptions of social media and misinformation. Seven said that the discussion about how to identify misinformation was especially important: \textit{“I guess it helped me to think about like how I detect misinformation”}, \textit{ “the nuances between how to tell if something is misinformation or fake news”}, \textit{“the most important is the ability to discriminate between false and fraud information”}, \textit{“it is very important for us to be able to divide which is false information and which was correct information and for me the most takeaway information here is how can we understand from other people's perspective about what such kind of information what to say”}. Five participants pointed to the discussion about the importance and effects of social media in everyday life, realizing that \textit{“it is very important to actually try and take time away from social media”} and the increasing influence social media has \textit{“when I was growing up there was no social media so it's really interesting to see how it became really like so important in our life”}, \textit{“not to spend much time with social media and look more for information from the books or interact directly with other people”}. Thinking about cognitive biases and how to deal with views other than their own was mentioned by only two participants.



\section{Summary and Discussion}
Our findings clearly show that social media is an important part of people's everyday life. 
Moreover, focus group discussions indicated that social media is not simply about connecting with family and friends; rather it has become a source of information, even though participants were conscious about the lack of reliability of this information. 

Misinformation was widely regarded as a common occurrence. The discussion revealed three main challenges that participants were struggling with: 1) the large amount of (mis)information, making it more difficult to filter any information; 2) problems with identifying misinformation, especially when facts and false information are mixed; and 3) the mainly negative emotional impact of misinformation on their own feelings but also their feelings towards others. Four patterns related to acting on misinformation also became apparent: 1) ignoring it, especially when deemed not personally relevant; 2) warning family and friends; 3) reporting it to the platform; and 4) cross-validation using other information sources.

Participants also experienced different levels of engagement when processing information on social media, ranging from low (e.g., ignoring), to medium (e.g., evaluating personal relevance, checking information about the author or information source, identifying keywords that could signal misinformation), to high (e.g., looking up the original information source, discussing with friends, family, or colleagues, searching for additional information about the same topic in other information sources). However, none of the participants had a rigorous strategy for identifying misinformation. Indeed, participants noted that the focus group discussion itself allowed them to understand how misinformation can be detected. 


A limitation of our study is the small sample (n = 16) which might not be representative of social media users in general. Furthermore, 
limiting the topic to mask usage might have influenced how participants responded in the discussions. Further studies on a variety of topics are necessary to validate our findings. 

\section{Implications}
Our findings provide insights into how social media users confront and react to information in general and misinformation specifically. Based on these, we have identified three areas where social media users may need support.

\subsection{Information Selection}
Social media users are confronted with a large volume of information coming from different sources and platforms. People struggle to filter the information flow and identify the truthfulness of the information. Intelligent tools could support users in two different ways. (1) \emph{Making misinformation less visible}. Misinformation can be completely removed (an approach already used by some providers), or it can be camouflaged with opaque overlays or patterns that make it less visible (for images) or less readable (for text). Many previous studies have looked at highlighting misinformation by adding warnings (e.g., \cite{Bailey2021, Mena2020, Ross2018}. However, as our findings indicate, people are overloaded with information. Additionally, salient cues like warnings might draw users' attention to misinformation rather than away from it. In general, warnings are designed to capture attention and hold it long enough for the user to process the content (e.g. \cite{Wogalter2021}). Camouflaging misinformation without removing it completely increases algorithmic transparency. Providing a way for the user to control the overlay obscuring the information ensures that they maintain autonomy. Importantly, content obfuscation and controls over it must be considered in light of disability. 
(2) \emph{Highlighting key information}. Highlighting information 
based on simple strategies to determine its truthfulness
might speed up 
recognition of the degree of truthfulness and ease anxieties about credible sources being infiltrated by misinformation. Such key information can be common phrases often used in misinformation ("\emph{share this quickly!!!}"), or easily accessible information about the information source (original source, author, evidence cited). Subtle visual emphasis and timed animations could draw attention to these elements, especially if the user has been attending to the information for some time. 

\subsection{Information Engagement}
Given the large amount of information that social media users are exposed to on a daily basis, it is understandable that many choose to ignore those that they do not consider relevant or that they deem misinformation. However, engaging with differing viewpoints can encourage a re-evaluation of one's beliefs and motivate users to think more deeply about an issue. Of course, this influence could go both ways: in opposition to misinformation or towards accepting misinformation as a legitimate alternative view. Indeed, only a couple of participants identified raising awareness of cognitive biases and learning how to handle opposing views as important takeaways. Encouraging users to engage more deeply with the information that they are exposed to could therefore trigger metacognitive processes to reconsider one's own position, reiterate pro and con arguments, and raise awareness of one's own thought processes. An intelligent tool could do this by asking thought-provoking questions, such as "Do you think this information is true or false?" "Why do you think so?" "What are the arguments against this position?". In addition, an intelligent tool could help users cross-reference information by providing links to other sources that contain factual information or references to discussion groups that address the issue, i.e., diversification of information, similar to diversification of authors and communication styles \cite{Engelmann2019}. The usefulness of the focus group format itself suggests that it might also be helpful to encourage people to discuss and cross-reference content with friends, family, or colleagues.

\subsection{Emotion Management}
Emotions play an important role in people’s judgments, information evaluation, and information selection, and thus also affect how we react to certain stimuli \cite{Dolan2002}. Previous research \cite{Ferrara2015, Kramer2014, Naskar2020} as well as our study show that encountering misinformation or conflicting viewpoints triggers a variety of emotional reactions, many with a negative valence. 
For example, emotion recognition systems could provide feedback to users about their emotional responses to information, helping them to better understand their own reactions and the influence of their emotions on thoughts and actions. They could also highlight in advance the sentiments expressed in content not yet read. Emotion monitoring systems could prompt users to take time out or focus more on self-care when extreme emotions are detected (e.g., anger, intense resentment, shock) or when negative emotions persist for an extended period of time. 

\section{Conclusion}
Misinformation is difficult to detect and emotionally reactive. Current platforms and tools, and arguably education and social environments, are not sufficient to help people make sense of the content they are exposed to online, especially if it is divisive. Yet, as our findings show, people do exercise metacognitive skills without training or prompting. Yet, these skills may not be sufficient to recognize and handle responses to all forms of negative social media content, or indeed all forms of misinformation, at all times. As we have suggested, these informal strategies may be translated into design guidelines for intelligent tools that 
take on the burden of detecting divisive content and assisting people during social media use. Such tools may help everyday social media users to develop the metacognitive skills that they need to handle the infodemic while retaining their autonomy and continuing to enjoy the brighter side of social media. 

\begin{acks}
This research was supported by DLab Challenge : Laboratory for Design of Social Innovation in Global Networks (DLab) Research Grant.

\end{acks}

\bibliographystyle{ACM-Reference-Format}
\bibliography{CitationsFocusGroup}


\begin{thebibliography}{52}


\ifx \showCODEN    \undefined \def \showCODEN     #1{\unskip}     \fi
\ifx \showDOI      \undefined \def \showDOI       #1{#1}\fi
\ifx \showISBNx    \undefined \def \showISBNx     #1{\unskip}     \fi
\ifx \showISBNxiii \undefined \def \showISBNxiii  #1{\unskip}     \fi
\ifx \showISSN     \undefined \def \showISSN      #1{\unskip}     \fi
\ifx \showLCCN     \undefined \def \showLCCN      #1{\unskip}     \fi
\ifx \shownote     \undefined \def \shownote      #1{#1}          \fi
\ifx \showarticletitle \undefined \def \showarticletitle #1{#1}   \fi
\ifx \showURL      \undefined \def \showURL       {\relax}        \fi
\providecommand\bibfield[2]{#2}
\providecommand\bibinfo[2]{#2}
\providecommand\natexlab[1]{#1}
\providecommand\showeprint[2][]{arXiv:#2}

\bibitem[\protect\citeauthoryear{Abd El-Jawad, Hodhod, and Omar}{Abd El-Jawad
  et~al\mbox{.}}{2018}]%
        {AbdElJawad2018}
\bibfield{author}{\bibinfo{person}{Mohammed~H Abd El-Jawad},
  \bibinfo{person}{Rania Hodhod}, {and} \bibinfo{person}{Yasser~MK Omar}.}
  \bibinfo{year}{2018}\natexlab{}.
\newblock \showarticletitle{Sentiment analysis of social media networks using
  machine learning}. In \bibinfo{booktitle}{\emph{2018 14th international
  computer engineering conference (ICENCO)}}. \bibinfo{publisher}{IEEE},
  \bibinfo{pages}{174--176}.
\newblock
\showISBNx{1538651173}


\bibitem[\protect\citeauthoryear{Alduaiji and Datta}{Alduaiji and
  Datta}{2018}]%
        {Alduaiji2018}
\bibfield{author}{\bibinfo{person}{Noha Alduaiji} {and}
  \bibinfo{person}{Amitava Datta}.} \bibinfo{year}{2018}\natexlab{}.
\newblock \showarticletitle{An empirical study on sentiments in twitter
  communities}. In \bibinfo{booktitle}{\emph{2018 IEEE International Conference
  on Data Mining Workshops (ICDMW)}}. \bibinfo{publisher}{IEEE},
  \bibinfo{pages}{1166--1172}.
\newblock
\showISBNx{1538692880}


\bibitem[\protect\citeauthoryear{Allcott, Gentzkow, and Yu}{Allcott
  et~al\mbox{.}}{2019}]%
        {Allcott2019}
\bibfield{author}{\bibinfo{person}{Hunt Allcott}, \bibinfo{person}{Matthew
  Gentzkow}, {and} \bibinfo{person}{Chuan Yu}.}
  \bibinfo{year}{2019}\natexlab{}.
\newblock \showarticletitle{Trends in the diffusion of misinformation on social
  media}.
\newblock \bibinfo{journal}{\emph{Research \& Politics}} \bibinfo{volume}{6},
  \bibinfo{number}{2} (\bibinfo{year}{2019}),
  \bibinfo{pages}{2053168019848554}.
\newblock
\showISSN{2053-1680}


\bibitem[\protect\citeauthoryear{Andalibi and Buss}{Andalibi and Buss}{2020}]%
        {Andalibi2020}
\bibfield{author}{\bibinfo{person}{Nazanin Andalibi} {and}
  \bibinfo{person}{Justin Buss}.} \bibinfo{year}{2020}\natexlab{}.
\newblock \showarticletitle{The human in emotion recognition on social media:
  Attitudes, outcomes, risks}. In \bibinfo{booktitle}{\emph{Proceedings of the
  2020 CHI Conference on Human Factors in Computing Systems}}.
  \bibinfo{pages}{1--16}.
\newblock


\bibitem[\protect\citeauthoryear{Andrews-Todd, Salovich, and Rapp}{Andrews-Todd
  et~al\mbox{.}}{2021}]%
        {AndrewsTodd2021}
\bibfield{author}{\bibinfo{person}{Jessica Andrews-Todd},
  \bibinfo{person}{Nikita~A Salovich}, {and} \bibinfo{person}{David~N Rapp}.}
  \bibinfo{year}{2021}\natexlab{}.
\newblock \showarticletitle{Differential effects of pressure on social
  contagion of memory}.
\newblock \bibinfo{journal}{\emph{Journal of Experimental Psychology: Applied}}
  (\bibinfo{year}{2021}).
\newblock
\showISSN{1939-2192}


\bibitem[\protect\citeauthoryear{Bailey, Olaguez, Klemfuss, and Loftus}{Bailey
  et~al\mbox{.}}{2021}]%
        {Bailey2021}
\bibfield{author}{\bibinfo{person}{Natasha~A Bailey}, \bibinfo{person}{Alma~P
  Olaguez}, \bibinfo{person}{Jessica~Zoe Klemfuss}, {and}
  \bibinfo{person}{Elizabeth~F Loftus}.} \bibinfo{year}{2021}\natexlab{}.
\newblock \showarticletitle{Tactics for increasing resistance to
  misinformation}.
\newblock \bibinfo{journal}{\emph{Applied Cognitive Psychology}}
  (\bibinfo{year}{2021}).
\newblock
\showISSN{0888-4080}


\bibitem[\protect\citeauthoryear{Bautista, Zhang, and Gwizdka}{Bautista
  et~al\mbox{.}}{2021}]%
        {Bautista2021}
\bibfield{author}{\bibinfo{person}{John~Robert Bautista}, \bibinfo{person}{Yan
  Zhang}, {and} \bibinfo{person}{Jacek Gwizdka}.}
  \bibinfo{year}{2021}\natexlab{}.
\newblock \showarticletitle{Healthcare professionals’ acts of correcting
  health misinformation on social media}.
\newblock \bibinfo{journal}{\emph{International Journal of Medical
  Informatics}}  \bibinfo{volume}{148} (\bibinfo{year}{2021}),
  \bibinfo{pages}{104375}.
\newblock
\showISSN{1386-5056}


\bibitem[\protect\citeauthoryear{Brashier, Eliseev, and Marsh}{Brashier
  et~al\mbox{.}}{2020}]%
        {Brashier2020}
\bibfield{author}{\bibinfo{person}{Nadia~M Brashier},
  \bibinfo{person}{Emmaline~Drew Eliseev}, {and} \bibinfo{person}{Elizabeth~J
  Marsh}.} \bibinfo{year}{2020}\natexlab{}.
\newblock \showarticletitle{An initial accuracy focus prevents illusory truth}.
\newblock \bibinfo{journal}{\emph{Cognition}}  \bibinfo{volume}{194}
  (\bibinfo{year}{2020}), \bibinfo{pages}{104054}.
\newblock
\showISSN{0010-0277}


\bibitem[\protect\citeauthoryear{Cuan-Baltazar, Muñoz-Perez, Robledo-Vega,
  Pérez-Zepeda, and Soto-Vega}{Cuan-Baltazar et~al\mbox{.}}{2020}]%
        {CuanBaltazar2020}
\bibfield{author}{\bibinfo{person}{Jose~Yunam Cuan-Baltazar},
  \bibinfo{person}{Maria~José Muñoz-Perez}, \bibinfo{person}{Carolina
  Robledo-Vega}, \bibinfo{person}{Maria~Fernanda Pérez-Zepeda}, {and}
  \bibinfo{person}{Elena Soto-Vega}.} \bibinfo{year}{2020}\natexlab{}.
\newblock \showarticletitle{Misinformation of COVID-19 on the internet:
  infodemiology study}.
\newblock \bibinfo{journal}{\emph{JMIR Public Health and Surveillance}}
  \bibinfo{volume}{6}, \bibinfo{number}{2} (\bibinfo{year}{2020}),
  \bibinfo{pages}{e18444}.
\newblock


\bibitem[\protect\citeauthoryear{Dolan}{Dolan}{2002}]%
        {Dolan2002}
\bibfield{author}{\bibinfo{person}{Raymond~J Dolan}.}
  \bibinfo{year}{2002}\natexlab{}.
\newblock \showarticletitle{Emotion, cognition, and behavior}.
\newblock \bibinfo{journal}{\emph{Science}} \bibinfo{volume}{298},
  \bibinfo{number}{5596} (\bibinfo{year}{2002}), \bibinfo{pages}{1191--1194}.
\newblock
\showISSN{0036-8075}


\bibitem[\protect\citeauthoryear{Engelmann, Kloss, Neuberger, and
  Brockmann}{Engelmann et~al\mbox{.}}{2019}]%
        {Engelmann2019}
\bibfield{author}{\bibinfo{person}{Ines Engelmann}, \bibinfo{person}{Andrea
  Kloss}, \bibinfo{person}{Christoph Neuberger}, {and} \bibinfo{person}{Tobias
  Brockmann}.} \bibinfo{year}{2019}\natexlab{}.
\newblock \showarticletitle{Visibility through information sharing: The role of
  tweet authors and communication styles in retweeting political information on
  Twitter}.
\newblock \bibinfo{journal}{\emph{International Journal of Communication}}
  \bibinfo{volume}{13} (\bibinfo{year}{2019}), \bibinfo{pages}{20}.
\newblock
\showISSN{1932-8036}


\bibitem[\protect\citeauthoryear{Ferrara and Yang}{Ferrara and Yang}{2015}]%
        {Ferrara2015}
\bibfield{author}{\bibinfo{person}{Emilio Ferrara} {and} \bibinfo{person}{Zeyao
  Yang}.} \bibinfo{year}{2015}\natexlab{}.
\newblock \showarticletitle{Measuring emotional contagion in social media}.
\newblock \bibinfo{journal}{\emph{PloS one}} \bibinfo{volume}{10},
  \bibinfo{number}{11} (\bibinfo{year}{2015}), \bibinfo{pages}{e0142390}.
\newblock
\showISSN{1932-6203}


\bibitem[\protect\citeauthoryear{Flavell}{Flavell}{1979}]%
        {Flavell1979}
\bibfield{author}{\bibinfo{person}{John~H Flavell}.}
  \bibinfo{year}{1979}\natexlab{}.
\newblock \showarticletitle{Metacognition and cognitive monitoring: A new area
  of cognitive–developmental inquiry}.
\newblock \bibinfo{journal}{\emph{American Psychologist}} \bibinfo{volume}{34},
  \bibinfo{number}{10} (\bibinfo{year}{1979}), \bibinfo{pages}{906}.
\newblock
\showISSN{1935-990X}


\bibitem[\protect\citeauthoryear{Flynn, Nyhan, and Reifler}{Flynn
  et~al\mbox{.}}{2017}]%
        {Flynn2017}
\bibfield{author}{\bibinfo{person}{DJ Flynn}, \bibinfo{person}{Brendan Nyhan},
  {and} \bibinfo{person}{Jason Reifler}.} \bibinfo{year}{2017}\natexlab{}.
\newblock \showarticletitle{The nature and origins of misperceptions:
  Understanding false and unsupported beliefs about politics}.
\newblock \bibinfo{journal}{\emph{Political Psychology}}  \bibinfo{volume}{38}
  (\bibinfo{year}{2017}), \bibinfo{pages}{127--150}.
\newblock
\showISSN{0162-895X}


\bibitem[\protect\citeauthoryear{Frenda, Nichols, and Loftus}{Frenda
  et~al\mbox{.}}{2011}]%
        {Frenda2011}
\bibfield{author}{\bibinfo{person}{Steven~J Frenda}, \bibinfo{person}{Rebecca~M
  Nichols}, {and} \bibinfo{person}{Elizabeth~F Loftus}.}
  \bibinfo{year}{2011}\natexlab{}.
\newblock \showarticletitle{Current issues and advances in misinformation
  research}.
\newblock \bibinfo{journal}{\emph{Current Directions in Psychological Science}}
  \bibinfo{volume}{20}, \bibinfo{number}{1} (\bibinfo{year}{2011}),
  \bibinfo{pages}{20--23}.
\newblock
\showISSN{0963-7214}


\bibitem[\protect\citeauthoryear{Geeng, Yee, and Roesner}{Geeng
  et~al\mbox{.}}{2020}]%
        {Geeng2020}
\bibfield{author}{\bibinfo{person}{Christine Geeng}, \bibinfo{person}{Savanna
  Yee}, {and} \bibinfo{person}{Franziska Roesner}.}
  \bibinfo{year}{2020}\natexlab{}.
\newblock \showarticletitle{Fake News on Facebook and Twitter: Investigating
  How People (Don't) Investigate}. In \bibinfo{booktitle}{\emph{Proceedings of
  the 2020 CHI Conference on Human Factors in Computing Systems}}.
  \bibinfo{pages}{1--14}.
\newblock


\bibitem[\protect\citeauthoryear{Guest, MacQueen, and Namey}{Guest
  et~al\mbox{.}}{2011}]%
        {Guest}
\bibfield{author}{\bibinfo{person}{Greg Guest}, \bibinfo{person}{Kathleen~M.
  MacQueen}, {and} \bibinfo{person}{Emily~E. Namey}.}
  \bibinfo{year}{2011}\natexlab{}.
\newblock \bibinfo{booktitle}{\emph{Applied Thematic Analysis}}.
\newblock \bibinfo{publisher}{Sage}.
\newblock


\bibitem[\protect\citeauthoryear{Hennessey}{Hennessey}{1999}]%
        {Hennessey1999}
\bibfield{author}{\bibinfo{person}{M~Gertrude Hennessey}.}
  \bibinfo{year}{1999}\natexlab{}.
\newblock \showarticletitle{Probing the Dimensions of Metacognition:
  Implications for Conceptual Change Teaching-Learning}.
\newblock  (\bibinfo{year}{1999}).
\newblock


\bibitem[\protect\citeauthoryear{Iturregui-Gallardo and
  Méndez-Ulrich}{Iturregui-Gallardo and Méndez-Ulrich}{2020}]%
        {IturreguiGallardo2020}
\bibfield{author}{\bibinfo{person}{Gonzalo Iturregui-Gallardo} {and}
  \bibinfo{person}{Jorge~Luis Méndez-Ulrich}.}
  \bibinfo{year}{2020}\natexlab{}.
\newblock \showarticletitle{Towards the creation of a tactile version of the
  Self-Assessment Manikin (T-SAM) for the emotional assessment of visually
  impaired people}.
\newblock \bibinfo{journal}{\emph{International Journal of Disability,
  Development and Education}} \bibinfo{volume}{67}, \bibinfo{number}{6}
  (\bibinfo{year}{2020}), \bibinfo{pages}{657--674}.
\newblock
\showISSN{1034-912X}


\bibitem[\protect\citeauthoryear{Kramer, Guillory, and Hancock}{Kramer
  et~al\mbox{.}}{2014}]%
        {Kramer2014}
\bibfield{author}{\bibinfo{person}{Adam~DI Kramer}, \bibinfo{person}{Jamie~E
  Guillory}, {and} \bibinfo{person}{Jeffrey~T Hancock}.}
  \bibinfo{year}{2014}\natexlab{}.
\newblock \showarticletitle{Experimental evidence of massive-scale emotional
  contagion through social networks}.
\newblock \bibinfo{journal}{\emph{Proceedings of the National Academy of
  Sciences}} \bibinfo{volume}{111}, \bibinfo{number}{24}
  (\bibinfo{year}{2014}), \bibinfo{pages}{8788--8790}.
\newblock
\showISSN{0027-8424}


\bibitem[\protect\citeauthoryear{Lambert and Lambert}{Lambert and
  Lambert}{2012}]%
        {Lambert2012}
\bibfield{author}{\bibinfo{person}{Vickie~A Lambert} {and}
  \bibinfo{person}{Clinton~E Lambert}.} \bibinfo{year}{2012}\natexlab{}.
\newblock \showarticletitle{Qualitative descriptive research: An acceptable
  design}.
\newblock \bibinfo{journal}{\emph{Pacific Rim International Journal of Nursing
  Research}} \bibinfo{volume}{16}, \bibinfo{number}{4} (\bibinfo{year}{2012}),
  \bibinfo{pages}{255--256}.
\newblock
\showISSN{2586-8373}


\bibitem[\protect\citeauthoryear{Lazarus}{Lazarus}{1982}]%
        {Lazarus1982}
\bibfield{author}{\bibinfo{person}{Richard~S Lazarus}.}
  \bibinfo{year}{1982}\natexlab{}.
\newblock \showarticletitle{Thoughts on the relations between emotion and
  cognition}.
\newblock \bibinfo{journal}{\emph{American Psychologist}} \bibinfo{volume}{37},
  \bibinfo{number}{9} (\bibinfo{year}{1982}), \bibinfo{pages}{1019}.
\newblock
\showISSN{1935-990X}


\bibitem[\protect\citeauthoryear{Lazer, Baum, Benkler, Berinsky, Greenhill,
  Menczer, Metzger, Nyhan, Pennycook, and Rothschild}{Lazer
  et~al\mbox{.}}{2018}]%
        {Lazer2018}
\bibfield{author}{\bibinfo{person}{David~MJ Lazer}, \bibinfo{person}{Matthew~A
  Baum}, \bibinfo{person}{Yochai Benkler}, \bibinfo{person}{Adam~J Berinsky},
  \bibinfo{person}{Kelly~M Greenhill}, \bibinfo{person}{Filippo Menczer},
  \bibinfo{person}{Miriam~J Metzger}, \bibinfo{person}{Brendan Nyhan},
  \bibinfo{person}{Gordon Pennycook}, {and} \bibinfo{person}{David
  Rothschild}.} \bibinfo{year}{2018}\natexlab{}.
\newblock \showarticletitle{The science of fake news}.
\newblock \bibinfo{journal}{\emph{Science}} \bibinfo{volume}{359},
  \bibinfo{number}{6380} (\bibinfo{year}{2018}), \bibinfo{pages}{1094--1096}.
\newblock
\showISSN{0036-8075}


\bibitem[\protect\citeauthoryear{Li, Lau, Mo, Su, Wu, Tang, and Qin}{Li
  et~al\mbox{.}}{2016}]%
        {Li2016}
\bibfield{author}{\bibinfo{person}{Jibin Li}, \bibinfo{person}{Joseph~TF Lau},
  \bibinfo{person}{Phoenix~KH Mo}, \bibinfo{person}{Xuefen Su},
  \bibinfo{person}{Anise~MS Wu}, \bibinfo{person}{Jie Tang}, {and}
  \bibinfo{person}{Zuguo Qin}.} \bibinfo{year}{2016}\natexlab{}.
\newblock \showarticletitle{Validation of the Social Networking Activity
  Intensity Scale among junior middle school students in China}.
\newblock \bibinfo{journal}{\emph{PLoS One}} \bibinfo{volume}{11},
  \bibinfo{number}{10} (\bibinfo{year}{2016}), \bibinfo{pages}{e0165695}.
\newblock
\showISSN{1932-6203}


\bibitem[\protect\citeauthoryear{Mena}{Mena}{2020}]%
        {Mena2020}
\bibfield{author}{\bibinfo{person}{Paul Mena}.}
  \bibinfo{year}{2020}\natexlab{}.
\newblock \showarticletitle{Cleaning up social media: The effect of warning
  labels on likelihood of sharing false news on Facebook}.
\newblock \bibinfo{journal}{\emph{Policy \& Internet}} \bibinfo{volume}{12},
  \bibinfo{number}{2} (\bibinfo{year}{2020}), \bibinfo{pages}{165--183}.
\newblock
\showISSN{1944-2866}


\bibitem[\protect\citeauthoryear{Miller}{Miller}{2000}]%
        {Miller2000}
\bibfield{author}{\bibinfo{person}{Earl~K Miller}.}
  \bibinfo{year}{2000}\natexlab{}.
\newblock \showarticletitle{The prefontral cortex and cognitive control}.
\newblock \bibinfo{journal}{\emph{Nature Reviews Neuroscience}}
  \bibinfo{volume}{1}, \bibinfo{number}{1} (\bibinfo{year}{2000}),
  \bibinfo{pages}{59--65}.
\newblock
\showISSN{1471-0048}


\bibitem[\protect\citeauthoryear{Mislove, Marcon, Gummadi, Druschel, and
  Bhattacharjee}{Mislove et~al\mbox{.}}{2007}]%
        {Mislove2007}
\bibfield{author}{\bibinfo{person}{Alan Mislove}, \bibinfo{person}{Massimiliano
  Marcon}, \bibinfo{person}{Krishna~P Gummadi}, \bibinfo{person}{Peter
  Druschel}, {and} \bibinfo{person}{Bobby Bhattacharjee}.}
  \bibinfo{year}{2007}\natexlab{}.
\newblock \showarticletitle{Measurement and analysis of online social
  networks}. In \bibinfo{booktitle}{\emph{Proceedings of the 7th ACM SIGCOMM
  Conference on Internet measurement}}. \bibinfo{pages}{29--42}.
\newblock


\bibitem[\protect\citeauthoryear{Naskar, Singh, Kumar, Nandi, and
  Rivaherrera}{Naskar et~al\mbox{.}}{2020}]%
        {Naskar2020}
\bibfield{author}{\bibinfo{person}{Debashis Naskar},
  \bibinfo{person}{Sanasam~Ranbir Singh}, \bibinfo{person}{Durgesh Kumar},
  \bibinfo{person}{Sukumar Nandi}, {and} \bibinfo{person}{Eva Onaindia de~la
  Rivaherrera}.} \bibinfo{year}{2020}\natexlab{}.
\newblock \showarticletitle{Emotion dynamics of public opinions on twitter}.
\newblock \bibinfo{journal}{\emph{ACM Transactions on Information Systems
  (TOIS)}} \bibinfo{volume}{38}, \bibinfo{number}{2} (\bibinfo{year}{2020}),
  \bibinfo{pages}{1--24}.
\newblock
\showISSN{1046-8188}


\bibitem[\protect\citeauthoryear{Nelson-Field, Riebe, and
  Newstead}{Nelson-Field et~al\mbox{.}}{2013}]%
        {NelsonField2013}
\bibfield{author}{\bibinfo{person}{Karen Nelson-Field}, \bibinfo{person}{Erica
  Riebe}, {and} \bibinfo{person}{Kellie Newstead}.}
  \bibinfo{year}{2013}\natexlab{}.
\newblock \showarticletitle{The emotions that drive viral video}.
\newblock \bibinfo{journal}{\emph{Australasian Marketing Journal}}
  \bibinfo{volume}{21}, \bibinfo{number}{4} (\bibinfo{year}{2013}),
  \bibinfo{pages}{205--211}.
\newblock
\showISSN{1839-3349}


\bibitem[\protect\citeauthoryear{Pessoa}{Pessoa}{2008}]%
        {Pessoa2008}
\bibfield{author}{\bibinfo{person}{Luiz Pessoa}.}
  \bibinfo{year}{2008}\natexlab{}.
\newblock \showarticletitle{On the relationship between emotion and cognition}.
\newblock \bibinfo{journal}{\emph{Nature Reviews Neuroscience}}
  \bibinfo{volume}{9}, \bibinfo{number}{2} (\bibinfo{year}{2008}),
  \bibinfo{pages}{148--158}.
\newblock
\showISSN{1471-0048}


\bibitem[\protect\citeauthoryear{Phelps}{Phelps}{2006}]%
        {Phelps2006}
\bibfield{author}{\bibinfo{person}{Elizabeth~A Phelps}.}
  \bibinfo{year}{2006}\natexlab{}.
\newblock \showarticletitle{Emotion and cognition: insights from studies of the
  human amygdala}.
\newblock \bibinfo{journal}{\emph{Annu. Rev. Psychol.}}  \bibinfo{volume}{57}
  (\bibinfo{year}{2006}), \bibinfo{pages}{27--53}.
\newblock
\showISSN{0066-4308}


\bibitem[\protect\citeauthoryear{Roozenbeek, Schneider, Dryhurst, Kerr,
  Freeman, Recchia, Van Der~Bles, and Van Der~Linden}{Roozenbeek
  et~al\mbox{.}}{2020}]%
        {Roozenbeek2020}
\bibfield{author}{\bibinfo{person}{Jon Roozenbeek}, \bibinfo{person}{Claudia~R
  Schneider}, \bibinfo{person}{Sarah Dryhurst}, \bibinfo{person}{John Kerr},
  \bibinfo{person}{Alexandra~LJ Freeman}, \bibinfo{person}{Gabriel Recchia},
  \bibinfo{person}{Anne~Marthe Van Der~Bles}, {and} \bibinfo{person}{Sander Van
  Der~Linden}.} \bibinfo{year}{2020}\natexlab{}.
\newblock \showarticletitle{Susceptibility to misinformation about COVID-19
  around the world}.
\newblock \bibinfo{journal}{\emph{Royal Society open science}}
  \bibinfo{volume}{7}, \bibinfo{number}{10} (\bibinfo{year}{2020}),
  \bibinfo{pages}{201199}.
\newblock
\showISSN{2054-5703}


\bibitem[\protect\citeauthoryear{Ross, Jung, Heisel, and Stieglitz}{Ross
  et~al\mbox{.}}{2018}]%
        {Ross2018}
\bibfield{author}{\bibinfo{person}{Björn Ross}, \bibinfo{person}{Anna Jung},
  \bibinfo{person}{Jennifer Heisel}, {and} \bibinfo{person}{Stefan Stieglitz}.}
  \bibinfo{year}{2018}\natexlab{}.
\newblock \showarticletitle{Fake news on social media: The (in) effectiveness
  of warning messages}.
\newblock  (\bibinfo{year}{2018}).
\newblock


\bibitem[\protect\citeauthoryear{Salovich and Rapp}{Salovich and Rapp}{2020}]%
        {Salovich2020}
\bibfield{author}{\bibinfo{person}{Nikita~A Salovich} {and}
  \bibinfo{person}{David~N Rapp}.} \bibinfo{year}{2020}\natexlab{}.
\newblock \showarticletitle{Misinformed and unaware? Metacognition and the
  influence of inaccurate information}.
\newblock \bibinfo{journal}{\emph{Journal of Experimental Psychology: Learning,
  Memory, and Cognition}} (\bibinfo{year}{2020}).
\newblock
\showISSN{1939-1285}


\bibitem[\protect\citeauthoryear{Saltz, Leibowicz, and Wardle}{Saltz
  et~al\mbox{.}}{2021}]%
        {Saltz2021}
\bibfield{author}{\bibinfo{person}{Emily Saltz}, \bibinfo{person}{Claire~R
  Leibowicz}, {and} \bibinfo{person}{Claire Wardle}.}
  \bibinfo{year}{2021}\natexlab{}.
\newblock \showarticletitle{Encounters with Visual Misinformation and Labels
  Across Platforms: An Interview and Diary Study to Inform Ecosystem Approaches
  to Misinformation Interventions}. In \bibinfo{booktitle}{\emph{Extended
  Abstracts of the 2021 CHI Conference on Human Factors in Computing Systems}}.
  \bibinfo{pages}{1--6}.
\newblock


\bibitem[\protect\citeauthoryear{Sandelowski}{Sandelowski}{2000}]%
        {Sandelowski2000}
\bibfield{author}{\bibinfo{person}{Margarete Sandelowski}.}
  \bibinfo{year}{2000}\natexlab{}.
\newblock \showarticletitle{Whatever happened to qualitative description?}
\newblock \bibinfo{journal}{\emph{Research in Nursing \& Health}}
  \bibinfo{volume}{23}, \bibinfo{number}{4} (\bibinfo{year}{2000}),
  \bibinfo{pages}{334--340}.
\newblock
\showISSN{0160-6891}


\bibitem[\protect\citeauthoryear{Schreiner, Fischer, and Riedl}{Schreiner
  et~al\mbox{.}}{2021}]%
        {Schreiner2021}
\bibfield{author}{\bibinfo{person}{Melanie Schreiner}, \bibinfo{person}{Thomas
  Fischer}, {and} \bibinfo{person}{Rene Riedl}.}
  \bibinfo{year}{2021}\natexlab{}.
\newblock \showarticletitle{Impact of content characteristics and emotion on
  behavioral engagement in social media: literature review and research
  agenda}.
\newblock \bibinfo{journal}{\emph{Electronic Commerce Research}}
  \bibinfo{volume}{21}, \bibinfo{number}{2} (\bibinfo{year}{2021}),
  \bibinfo{pages}{329--345}.
\newblock
\showISSN{1572-9362}


\bibitem[\protect\citeauthoryear{Shin, Jian, Driscoll, and Bar}{Shin
  et~al\mbox{.}}{2018}]%
        {Shin2018}
\bibfield{author}{\bibinfo{person}{Jieun Shin}, \bibinfo{person}{Lian Jian},
  \bibinfo{person}{Kevin Driscoll}, {and} \bibinfo{person}{François Bar}.}
  \bibinfo{year}{2018}\natexlab{}.
\newblock \showarticletitle{The diffusion of misinformation on social media:
  Temporal pattern, message, and source}.
\newblock \bibinfo{journal}{\emph{Computers in Human Behavior}}
  \bibinfo{volume}{83} (\bibinfo{year}{2018}), \bibinfo{pages}{278--287}.
\newblock
\showISSN{0747-5632}


\bibitem[\protect\citeauthoryear{Sternberg}{Sternberg}{1986a}]%
        {Sternberg1986a}
\bibfield{author}{\bibinfo{person}{Robert~J Sternberg}.}
  \bibinfo{year}{1986}\natexlab{a}.
\newblock \showarticletitle{Inside Intelligence: Cognitive science enables us
  to go beyond intelligence tests and understand how the human mind solves
  problems}.
\newblock \bibinfo{journal}{\emph{American Scientist}} \bibinfo{volume}{74},
  \bibinfo{number}{2} (\bibinfo{year}{1986}), \bibinfo{pages}{137--143}.
\newblock
\showISSN{0003-0996}


\bibitem[\protect\citeauthoryear{Sternberg}{Sternberg}{1986b}]%
        {Sternbergb}
\bibfield{author}{\bibinfo{person}{Robert~J Sternberg}.}
  \bibinfo{year}{1986}\natexlab{b}.
\newblock \bibinfo{booktitle}{\emph{A triarchic theory of human intelligence}}.
\newblock \bibinfo{publisher}{Springer}, \bibinfo{pages}{43--44}.
\newblock


\bibitem[\protect\citeauthoryear{Tasnim, Hossain, and Mazumder}{Tasnim
  et~al\mbox{.}}{2020}]%
        {Tasnim2020}
\bibfield{author}{\bibinfo{person}{Samia Tasnim}, \bibinfo{person}{Md~Mahbub
  Hossain}, {and} \bibinfo{person}{Hoimonty Mazumder}.}
  \bibinfo{year}{2020}\natexlab{}.
\newblock \showarticletitle{Impact of rumors and misinformation on COVID-19 in
  social media}.
\newblock \bibinfo{journal}{\emph{Journal of Preventive Medicine and Public
  Health}} \bibinfo{volume}{53}, \bibinfo{number}{3} (\bibinfo{year}{2020}),
  \bibinfo{pages}{171--174}.
\newblock
\showISSN{1975-8375}


\bibitem[\protect\citeauthoryear{Vashishtha and Susan}{Vashishtha and
  Susan}{2019}]%
        {Vashishtha2019}
\bibfield{author}{\bibinfo{person}{Srishti Vashishtha} {and}
  \bibinfo{person}{Seba Susan}.} \bibinfo{year}{2019}\natexlab{}.
\newblock \showarticletitle{Fuzzy rule based unsupervised sentiment analysis
  from social media posts}.
\newblock \bibinfo{journal}{\emph{Expert Systems with Applications}}
  \bibinfo{volume}{138} (\bibinfo{year}{2019}), \bibinfo{pages}{112834}.
\newblock
\showISSN{0957-4174}


\bibitem[\protect\citeauthoryear{Venkatesh}{Venkatesh}{2000}]%
        {Venkatesh2000}
\bibfield{author}{\bibinfo{person}{Viswanath Venkatesh}.}
  \bibinfo{year}{2000}\natexlab{}.
\newblock \showarticletitle{Determinants of perceived ease of use: Integrating
  control, intrinsic motivation, and emotion into the technology acceptance
  model}.
\newblock \bibinfo{journal}{\emph{Information Systems Research}}
  \bibinfo{volume}{11}, \bibinfo{number}{4} (\bibinfo{year}{2000}),
  \bibinfo{pages}{342--365}.
\newblock
\showISSN{1047-7047}


\bibitem[\protect\citeauthoryear{Vraga and Bode}{Vraga and Bode}{2017}]%
        {Vraga2017}
\bibfield{author}{\bibinfo{person}{Emily~K Vraga} {and}
  \bibinfo{person}{Leticia Bode}.} \bibinfo{year}{2017}\natexlab{}.
\newblock \showarticletitle{Using expert sources to correct health
  misinformation in social media}.
\newblock \bibinfo{journal}{\emph{Science Communication}} \bibinfo{volume}{39},
  \bibinfo{number}{5} (\bibinfo{year}{2017}), \bibinfo{pages}{621--645}.
\newblock
\showISSN{1075-5470}


\bibitem[\protect\citeauthoryear{Walsh, Fielder, Carey, and Carey}{Walsh
  et~al\mbox{.}}{2013}]%
        {Walsh2013}
\bibfield{author}{\bibinfo{person}{Jennifer~L Walsh}, \bibinfo{person}{Robyn~L
  Fielder}, \bibinfo{person}{Kate~B Carey}, {and} \bibinfo{person}{Michael~P
  Carey}.} \bibinfo{year}{2013}\natexlab{}.
\newblock \showarticletitle{Female college students’ media use and academic
  outcomes: Results from a longitudinal cohort study}.
\newblock \bibinfo{journal}{\emph{Emerging Adulthood}} \bibinfo{volume}{1},
  \bibinfo{number}{3} (\bibinfo{year}{2013}), \bibinfo{pages}{219--232}.
\newblock
\showISSN{2167-6968}


\bibitem[\protect\citeauthoryear{Walter, Brooks, Saucier, and Suresh}{Walter
  et~al\mbox{.}}{2020}]%
        {Walter2020}
\bibfield{author}{\bibinfo{person}{Nathan Walter}, \bibinfo{person}{John~J
  Brooks}, \bibinfo{person}{Camille~J Saucier}, {and} \bibinfo{person}{Sapna
  Suresh}.} \bibinfo{year}{2020}\natexlab{}.
\newblock \showarticletitle{Evaluating the impact of attempts to correct health
  misinformation on social media: A meta-analysis}.
\newblock \bibinfo{journal}{\emph{Health Communication}}
  (\bibinfo{year}{2020}), \bibinfo{pages}{1--9}.
\newblock
\showISSN{1041-0236}


\bibitem[\protect\citeauthoryear{Wang, He, Xu, and Zhang}{Wang
  et~al\mbox{.}}{2020}]%
        {Wang2020}
\bibfield{author}{\bibinfo{person}{Rui Wang}, \bibinfo{person}{Yuan He},
  \bibinfo{person}{Jing Xu}, {and} \bibinfo{person}{Hongzhong Zhang}.}
  \bibinfo{year}{2020}\natexlab{}.
\newblock \showarticletitle{Fake news or bad news? Toward an emotion-driven
  cognitive dissonance model of misinformation diffusion}.
\newblock \bibinfo{journal}{\emph{Asian Journal of Communication}}
  \bibinfo{volume}{30}, \bibinfo{number}{5} (\bibinfo{year}{2020}),
  \bibinfo{pages}{317--342}.
\newblock
\showISSN{0129-2986}


\bibitem[\protect\citeauthoryear{Wogalter, Mayhorn, and Laughery~Sr}{Wogalter
  et~al\mbox{.}}{2021}]%
        {Wogalter2021}
\bibfield{author}{\bibinfo{person}{Michael~S Wogalter},
  \bibinfo{person}{Christopher~B Mayhorn}, {and} \bibinfo{person}{Kenneth~R
  Laughery~Sr}.} \bibinfo{year}{2021}\natexlab{}.
\newblock \showarticletitle{Warnings and hazard communications}.
\newblock \bibinfo{journal}{\emph{Handbook of human factors and ergonomics}}
  (\bibinfo{year}{2021}), \bibinfo{pages}{644--667}.
\newblock


\bibitem[\protect\citeauthoryear{Wu, Morstatter, Hu, and Liu}{Wu
  et~al\mbox{.}}{2016}]%
        {Wu2016}
\bibfield{author}{\bibinfo{person}{Liang Wu}, \bibinfo{person}{Fred
  Morstatter}, \bibinfo{person}{Xia Hu}, {and} \bibinfo{person}{Huan Liu}.}
  \bibinfo{year}{2016}\natexlab{}.
\newblock \bibinfo{booktitle}{\emph{Mining misinformation in social media}}.
\newblock \bibinfo{publisher}{Chapman and Hall/CRC}, \bibinfo{pages}{135--162}.
\newblock
\showISBNx{131539670X}


\bibitem[\protect\citeauthoryear{Yoo, Song, and Jeong}{Yoo
  et~al\mbox{.}}{2018}]%
        {Yoo2018}
\bibfield{author}{\bibinfo{person}{SoYeop Yoo}, \bibinfo{person}{JeIn Song},
  {and} \bibinfo{person}{OkRan Jeong}.} \bibinfo{year}{2018}\natexlab{}.
\newblock \showarticletitle{Social media contents based sentiment analysis and
  prediction system}.
\newblock \bibinfo{journal}{\emph{Expert Systems with Applications}}
  \bibinfo{volume}{105} (\bibinfo{year}{2018}), \bibinfo{pages}{102--111}.
\newblock
\showISSN{0957-4174}


\bibitem[\protect\citeauthoryear{Yu}{Yu}{2014}]%
        {Yu2014}
\bibfield{author}{\bibinfo{person}{Jusheng Yu}.}
  \bibinfo{year}{2014}\natexlab{}.
\newblock \showarticletitle{We look for social, not promotion: Brand post
  strategy, consumer emotions, and engagement}.
\newblock \bibinfo{journal}{\emph{International Journal of Media \&
  Communication}} \bibinfo{volume}{1}, \bibinfo{number}{2}
  (\bibinfo{year}{2014}), \bibinfo{pages}{28--37}.
\newblock


\bibitem[\protect\citeauthoryear{Zajonc}{Zajonc}{1984}]%
        {Zajonc1984}
\bibfield{author}{\bibinfo{person}{Robert~B Zajonc}.}
  \bibinfo{year}{1984}\natexlab{}.
\newblock \showarticletitle{On the primacy of affect}.
\newblock  (\bibinfo{year}{1984}).
\newblock
\showISSN{1935-990X}


\end{thebibliography}

\end{document}